\documentclass[aps,preprint,showpacs,preprintnumbers,amsmath,amssymb]{revtex4}
\usepackage{graphicx}
\usepackage[all]{xy}

\begin{document}

\title{Point-Coupling Models from Mesonic Hypermassive Limit 
and Mean-Field Approaches}  
\author{O. Louren\c co$^{1,2}$, M. Dutra$^{1,2}$, R. L. P. G. Amaral$^{1,3}$,
and A. Delfino$^1$}

\affiliation{$^1$Instituto de F\'{\i}sica - Universidade Federal Fluminense,
Av. Litor\^ anea s/n, 24210-150 Boa Viagem, Niter\'oi RJ, Brazil\\
$^2$Departamento de F\'isica, Instituto Tecnol\'ogico da Aeron\'autica, CTA,
S\~ao Jos\'e dos Campos, 12228-900, SP, Brazil\\
 $^3$ Center for Theoretical Physics, Massachusetts Institute of Technology, Cambridge, MA 02138, USA}
\received{}
\revised{}
\date{\today} 

\begin{abstract}
In this work we show how nonlinear point-coupling models, described by a
Lagrangian density that presents only terms up to fourth order in the fermion 
condensate $(\bar{\psi}\psi)$, are derived from a modified meson-exchange 
nonlinear Walecka model. The derivation can be done through two distinct 
methods, namely, the hypermassive meson limit within a functional integral 
approach, and the mean-field approximation in which equations of state at zero
temperature of the nonlinear point-coupling models are directly obtained.
\end{abstract}
\pacs{ }

\maketitle

\section{INTRODUCTION}

The point-coupling interaction problem was first addressed in the early thirties
by L. H. Thomas \cite{thomas} investigating the range of the two-nucleon force.
As a side remark of his work, he observed that when the range of the two-body
force goes to zero with the two-body binding energy kept fixed the binding
energy of the quantum-three-body state goes to minus infinity. Decades later, a
new and apparently not related three-body effect was proposed by Efimov
\cite{efimov}. When a quantum two-body system has a zero energy bound state then
the three-body system will have an infinite number of bound states with an
accumulation point at the common two and three body threshold. Both, the Efimov
and the Thomas effects are universal, since the associated three-body wave
functions have long tails in the classically forbidden region outside the range
of the potential. In a unified momentum space description, based on ideas of
Amado and Noble \cite{amado}, it has been claimed that these two apparently
different effects are related to the same singular structure of the kernel of
the Faddeev equation \cite{delfino1}. On the other hand, in appropriate units,
the presence of one of these effects implies the presence of the other
\cite{delfino1,few}. The Thomas-Efimov effect explains very well some few-body
correlations \cite{tjon} and is conjectured to be behind the Coester band
\cite{coester} for different nuclear matter models \cite{delfino2}. 

Since relativistic hadronic point-coupling models, that have been used in
the description of infinite nuclear matter, and finite nuclei as well
\cite{nikolaus}, can be viewed as a connection between the well established
finite range relativistic models and the Skyrme ones \cite{nikolaus}, a better 
understanding of their structure becomes of interest when an important 
theoretical challenge is to construct a universal nuclear effective density 
functional \cite{vretenar}. 

In this work we deal with a specific nonlinear point-coupling model (NLPC)
described by a Fermionic Lagrangian density with interaction terms in third
and fourth powers of the scalar density operator, that we have used in Ref.
\cite{rubens} in a comparative study with the standard nonlinear Walecka
model(s), and in Ref. \cite{nrl}, in which after taking its nonrelativistic
limit, a generalized Skyrme energy density functional was obtained. 

We focus on the derivation of the NLPC model from finite range ones in two 
different ways. First, we present the infinitely massive meson limit within the
formal point of view of the integral functional approach. This method shows in a
clear fashion the equivalence of the usual Walecka model to the linear
point-coupling one. However, contrary to this case, NLW models are thereby not
formally equivalent to NLPC ones. Therefore, we pose the question on how
NLPC models can be derived if one insists in a meson exchange model as their
origin. To answer this question, we construct a modified nonlinear Walecka
(MNLW) model in which the limit of infinite meson masses leads exactly to
the Lagrangian density of the NLPC model. This MNLW model includes terms of
third and fourth powers of the scalar meson field, together with terms in lower
powers of the fields coupled to the Fermionic scalar density operator.
Preliminary results on the use of this hypermassive meson limit have been
presented in Ref. \cite{ijmp}. 

The traditional mean-field approximation, implemented with some physical 
requirements, is an alternative way to construct the NLPC models from the MNLW ones. 
In this case we show that the equations of state of the MNLW models, relating 
energy, density and pressure, are exactly the same of the NLPC ones.

We can thus synthesize the study of the NLW, MNLW, and NLPC models through the 
following diagram

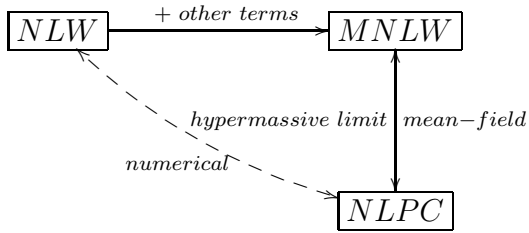
\vspace{0.5cm}
\xymatrix{
&&&&*+[F]{NLW} \ar@/_0.4cm/@{<-->}[rrrdd]_{numerical\,} 
\ar[rrr]^{\,\,+\,\,other\,\,terms\,\,}  & & & *+[F]{MNLW} \ar@{<->}[dd]^{mean-field}
_{hypermassive\,\,limit}\\ 
&&&&  & & &\\
&&&&  & & & *+[F]{NLPC}
}
\vspace{0.7cm}
\hspace{-0.6cm}where the numerical equivalences between NLW and NLPC models have 
been analyzed in Ref. \cite{rubens}, and the different connections among MNLW 
and NLPC models will be studied in this work at an analytical level.

This work extends the study presented in \cite{rubens,ijmp} through the
following points:

$\bullet$ A detailed and rigorous derivation of the point-coupling models 
from the modified NLW ones is done by using the hypermassive meson limit in the 
functional integral method. From this approach we show how the linear PC models 
can be obtained from the Walecka one(s), and in the same way, how the MNLW
model(s) generates the NLPC one(s).

$\bullet$ The mean-field approach in the no-sea approximation is also used to 
construct the equations of state (EOS) of infinite nuclear matter of the
MNLW model. In this approximation we show that these EOS are exactly the same
of the NLPC ones.

Our paper is organized as follows. In Sec. II by using a functional integral 
formalism we derive the linear point-coupling model from the Walecka one. The
same study is extended to obtain the NLPC models from the MNLW ones. In Sec. III
we explicitly derive the equations of state of the MNLW model. Finally, the main
conclusions are summarized.

\section{HYPERMASSIVE MESON LIMIT} 

\subsection{Linear point-coupling model from the Walecka one}

We start with the Walecka model Lagrangian density given by \cite{walecka}
\begin{eqnarray}
\mathcal{L}_W &=& \bar{\psi}(i\gamma^\mu\partial_{\mu}-M)\psi 
+ \dfrac{1}{2}\partial^{\mu}\phi\partial_{\mu}\phi - \dfrac{1}{2}m_s^2\phi^2 
-  \dfrac{1}{4}F^{\mu\nu}F_{\mu\nu} + \dfrac{1}{2}m_V^2 V_{\mu} V^{\mu} 
- g_s \bar{\psi}\phi\psi \nonumber \\ 
&-& g_V \bar{\psi}\gamma^{\mu}V_{\mu}\psi 
\label{dlwalecka}
\end{eqnarray}
where $F^{\mu\nu}\equiv\partial^{\mu}V^{\nu}-\partial^{\nu}V^{\mu}\,$.

In this Lagrangian, $\psi$, $\phi$ and $V^{\mu}$ are the nucleon, scalar and 
vector fields respectively and $M$, $m_{s}$ and $m_{v}$ refer to bare nucleon 
and the mesonic $\sigma$ and $\omega$  masses respectively.
 
The fermionic, scalar and vector fields constitute an irreducible set of 
generators  $\{\psi,\bar\psi, \phi, V_\mu\}$ of the intrinsic local algebra of
fields of the model, ${\bf A}$. The polynomial algebra of intrinsic fields
allows for the construction of the net of Wightman functions. From the Wightman
functions of the polynomial algebra of intrinsic fields the physical Hilbert
space is reconstructed, ${\mathbf{H}}=\mathbf{A}\left| 0\right\rangle $, thus
defining the physical content of the model \cite{morchio}. Under this setting an
equivalence of models should be understood as an assertion on the isomorphism of
their physical Hilbert spaces. This kind of equivalence is extremely rare to
occur. It is believed to occur in the context of duality transformation and is
witnessed in the context of two-dimensional field theory as in the bosonization
phenomenon \cite{belvedere1}. In establishing this isomorphism, operator
and functional integral methods are usually complementary tools \cite{books}.

Our concern here, however, is with much less stringent equivalences. We shall, 
first, derive an equivalence of the usual Walecka model to the linear
point-coupling model with terms of second order in the fermionic density and 
vector current (fourth order in the fermion fields). This is not an equivalence 
of the Hilbert spaces  but of the physical content of the models amenable to 
mean-field procedures. The mean-field  procedures start from discarding the 
kinetic terms for the scalar and vector fields what is related to the infinite 
limit of the masses of the bosons. The derivation of this equivalence will be 
provided here using functional integral methods. The presentation of this 
treatment at length in this section for this case is motivated by the need of 
clarifying the ideas to be used later in the case of interaction Lagrangian
models, which contains higher power of the mesonic fields. It will be thus
distinguished what is and what is not valid for these more involved models. 

We start by constructing the generating functionals within the functional
integral formalism from which the correlation functions are obtained. We will
then use this formalism to connect the Walecka model to the linear
point-coupling model. 

For the Walecka model, the generating functional is given by 
\begin{equation}
W[J,A_{\mu},\eta,\bar{\eta}]=  N \int[D\psi][D\bar{\psi}][DV^{\mu}][D\phi]
\displaystyle\ \mbox{e}^{iS_S}
\label{fpwalecka}
\end{equation}
with  
\begin{eqnarray}
N^{-1} &=&  {\displaystyle \int d^4x\ \mbox{e}^{iS} }\,\mbox{,} \\
S   &=& \int d^4x\ \mathcal{L}_W \quad \mbox{and}
\label{acaowalecka} \\
S_S &=& \int d^4x\ \left[\mathcal{L}_W + A_{\mu}(x)V^{\mu}(x) + J(x)\phi(x) 
+ \bar{\eta}(x)\psi(x) + \eta(x)\bar{\psi}(x)\right] 
\label{acaowaleckafonte}
\end{eqnarray}
where  $A_{\mu}(x)$, $J(x)$, $\eta(x)$ e $\bar{\eta}(x)$ are the sources for
vectorial ($V^{\mu}$), scalar  ($\phi$) and spinorial  ($\bar{\psi}$ and $\psi$)
fields respectively. $S_S$ and $S$ are the actions with and without the source
terms. Here it will be important to make the definitions 
$ V'^{\mu} \equiv m_V V^{\mu} $, $\phi ' \equiv m_s\phi\,$,
$ G'_s \equiv g_{s}/m_{s}$, $G'_V \equiv g_{V}/m_{V}$ , 
$A'_{\mu} \equiv A_{\mu}/m_{V}$ and $J' \equiv J/m_s $. 
With these definitions we can write 
\begin{eqnarray}
\mathcal{L}_W = \mathcal{L}'_W + \dfrac{1}{2m_s^2}\partial^{\mu}\phi'
\partial_{\mu}\phi' -  \dfrac{1}{4m_V^2}F'^{\mu\nu}F'_{\mu\nu} 
&\equiv& \mathcal{L}'_W + U(\phi',V'^\mu)
\end{eqnarray}
where 
\begin{eqnarray}
F'^{\mu\nu} &=& \partial^{\mu}V'^{\nu}-\partial^{\nu}V'^{\mu} \quad \mbox{and} \\
\mathcal{L}'_W &=& \bar{\psi}(i\gamma^\mu\partial_{\mu}-M)\psi 
- \dfrac{1}{2}\phi'^2  + \dfrac{1}{2}V'_{\mu} V'^{\mu}- G'_s \bar{\psi}\phi'\psi 
- G'_V \bar{\psi}\gamma^{\mu}V'_{\mu}\psi\,\mbox{.}
\label{dlwaleckam2}
\end{eqnarray}

Now, the generating functional Eq. (\ref{fpwalecka}) may be written in the 
following form, 
\begin{equation}
W[J',A'_{\mu},\eta,\bar{\eta}]=N \int[D\psi][D\bar{\psi}][DV'^{\mu}][D\phi']
\displaystyle\ \mbox{e}^{\displaystyle i\left[\int d^4x\ U(\phi',V'^\mu) + S'_S\right]}
\label{fpwaleckam}
\end{equation}
where 
\begin{equation}
S'_S = \int d^4x\ \left[\mathcal{L}'_W + A'_{\mu}(x)V'^{\mu}(x) + J'(x)\phi'(x) 
+ \bar{\eta}(x)\psi(x) + \eta(x)\bar{\psi}(x)\right] \ \mbox{.}
\label{acaowaleckafontem}
\end{equation}
In order to make contact with mean-field methods we consider the limit in which 
the mesonic masses become very large, allowing that terms involving 
$1/m_{s}^{2}$ and $1/m_{V}^{2}$ may be treated perturbatively in a generating 
functional expansion. It is in this perspective that we identify the fields in 
 $U(\phi',V'^\mu)$ with the respective functional derivatives, 
\begin{eqnarray}
U(\phi',V'^\mu) &=& \dfrac{1}{2m_s^2}\partial^{\mu}\phi'\partial_{\mu}\phi' 
-  \dfrac{1}{4m_V^2}F'^{\mu\nu}F'_{\mu\nu} 
\label{cinetico} \\
&=& \frac{1}{2m_s^2}\left(\partial^{\mu}\frac{\delta}{\delta J'}\right)^2 
- \frac{1}{4m_V^2} \left(\partial^{\nu}\frac{\delta}{\delta A'_{\mu}}
- \partial^{\mu}\frac{\delta}{\delta A'_{\nu}}\right)^2  \nonumber \\
&=& U\left(\dfrac{\delta}{\delta J'},\dfrac{\delta}{\delta A'_\mu}\right) 
\mbox{,} \nonumber
\end{eqnarray}
which allows us to write (\ref{fpwaleckam}) as 
\begin{equation}
W[J',A'_{\mu},\eta,\bar{\eta}]=N\ \mbox{e}^{\displaystyle i\left[\int d^4x\ 
U\left(\dfrac{\delta}{\delta J'},\dfrac{\delta}{\delta A'_\mu}\right)\right]} 
\int[D\psi][D\bar{\psi}][DV'^{\mu}][D\phi']\ \mbox{e}^{iS'_S} \mbox{.}
\end{equation}

Up to now we have not changed the physical content of the Walecka model but 
merely rewritten the generating functional in a form suited to a non standard 
expansion. We consider in the following the zero order term of the expansion of 
the generating functional by using 
\begin{equation}
\mbox{e}^{\displaystyle i\left[\int d^4x\ U\left(\dfrac{\delta}{\delta J'},
\dfrac{\delta}{\delta A'_\mu}\right)\right]} \simeq 1
,\label{aproximacao}
\end{equation}
so that 
\begin{equation}
W_{MF}[J',A'_{\mu},\eta,\bar{\eta}]=N \int[D\psi][D\bar{\psi}][DV'^{\mu}][D\phi']
\ \mbox{e}^{iS'_S}
\label{fpwaleckam2}
\end{equation}
in which the kinetic terms associated to the mesonic fields are 
neglected and $W_{MF}$ refers to the generating functional associated to 
mean-field treatment.  

It is important to analyze the situation once again from the view point of the 
structural properties of the model. In the start the intrinsic algebra of fields
was generated by the irreducible set of fields 
${\mathbf{S_1}}=\{\psi, \bar\psi, \phi, V^\mu\}$. Now, in the zero order 
approximation, since the mesonic kinetic terms have been suppressed, the 
equations of motion will allow to express, in this case explicitly, the mesonic 
fields in terms of the fermion densities. The algebra of fields has been turned 
on a reducible algebra and the irreducible algebra of fields is now constructed 
from the fermionic fields only ${\mathbf{ S_2}}=\{\bar\psi, \psi\}$. This seems 
to be a mathematical counterpart of the spirit of mean-field treatment. Of 
course the mesonic fields are still in play in the dynamics of the model, since 
they are still coupled to the fermionic fields, but we have lost control on
the independent degrees of freedom associated to the mesonic fields. 
The physical picture associated to this mathematical aspect will
be discussed at the end of this and the next subsections.

We proceed to the last step of the process, expressing the mesonic fields in 
terms of the fermion degrees and expressing the dynamics solely in terms of 
fermion fields. We do this in the functional integral formalism by decoupling 
the mesonic fields reducing them to auxiliary fields devoid of physical content.
To decouple them, we will proceed with a transformation of the mesonic
fields to the auxiliary ones. Before this, since we are not interested in
analyzing the mesonic correlation functions 
we will do   $J'(x)=A'_\mu(x)=0$ in Eq. (\ref{acaowaleckafontem}). 
With this condition and the following identities,
\begin{eqnarray}
-\dfrac{1}{2}\phi'^2-G'_s\bar{\psi}\phi'\psi &=& -\dfrac{1}{2}(\phi'+G'_s 
\bar{\psi}\psi)^2 +\dfrac{1}{2}{G'_s}^2(\bar{\psi}\psi)^2  
\label{ident1}\\
\dfrac{1}{2}V'^\mu V'_\mu - G'_V \bar{\psi}\gamma^{\mu}V'_\mu \psi &=& 
\dfrac{1}{2}(V'^\mu - G'_V \bar{\psi}\gamma^{\mu}\psi)^2 
- \dfrac{1}{2}{G'_V}^2(\bar{\psi}\gamma^{\mu}\psi)^2 \ \mbox{,}
\label{ident2}
\end{eqnarray}
Eq. (\ref{acaowaleckafontem}) may be rewritten as 
\begin{eqnarray}\label{sfprime}
S'_S &=& \int d^4x\ \left[\bar{\psi}(i\gamma^\mu\partial_{\mu}-M)\psi 
-\dfrac{1}{2}(\phi'+G'_s \bar{\psi}\psi)^2 + \dfrac{1}{2}{G'_s}^2
(\bar{\psi}\psi)^2 \right.
\nonumber \\ 
&+& \left. \dfrac{1}{2}(V'^\mu - G'_V \bar{\psi}\gamma^{\mu}\psi)^2 
- \dfrac{1}{2}{G'_V}^2(\bar{\psi}\gamma^{\mu}\psi)^2 + \bar{\eta}(x)\psi(x) 
+ \eta(x)\bar{\psi}(x) \right] \mbox{.}
\end{eqnarray}

Now, we define the change from the variables $\phi'$ and $V'^\mu$ to $\lambda$ 
and $R^\mu$
\begin{eqnarray}
\lambda &=& \phi'+G'_s \bar{\psi}\psi \\
R^\mu &=&  V'^\mu - G'_V \bar{\psi}\gamma^{\mu}\psi\,\mbox{.}
\end{eqnarray}

With this the generating functional, Eq. (\ref{fpwaleckam2}), becomes 
\begin{eqnarray}
W[\eta,\bar{\eta}]&=&{N}\int[D\lambda]\ \mbox{e}^{\displaystyle -i\int d^4x \ 
\dfrac{\lambda^2}{2}}
\int[DR^\mu]\ \mbox{e}^{\displaystyle i \int d^4x \dfrac{1}{2}R^\mu R_\mu} 
\int[D\psi][D\bar{\psi}]\ 
\mbox{e}^{i S''_S},\label{auxiliary}\\
&=&\mathcal{N} \int[D\psi][D\bar{\psi}]\ 
\mbox{e}^{i S''_S},
\end{eqnarray}
where 
\begin{eqnarray}
{\mathcal{N}^{-1}}&=& \int [D\psi][D\bar{\psi}] {\mbox{e}}^{i S''}  \\
S'' &=& \int d^4x\ \left[\bar{\psi}(i\gamma^\mu\partial_{\mu}-M)\psi 
+\dfrac{1}{2}{G'_s}^2(\bar{\psi}\psi)^2 - \dfrac{1}{2}{G'_V}^2(\bar{\psi}
\gamma^{\mu}\psi)^2 \right] \quad \mbox{and} 
\label{acaocontato2}\\
S''_S &=& \int d^4x \ \left[\bar{\psi}(i\gamma^\mu\partial_{\mu}-M)\psi 
+ \dfrac{1}{2}{G'_s}^2(\bar{\psi}\psi)^2 - \dfrac{1}{2}{G'_V}^2(\bar{\psi}
\gamma^{\mu}\psi)^2 + \bar{\eta}\psi + \eta\bar{\psi} \right] \ \mbox{.}
\label{acaofontecontato2}
\end{eqnarray}
Now, the Lagrangian density associated to 
Eqs. (\ref{acaocontato2})-(\ref{acaofontecontato2}) is given simply by 
\begin{equation}
\mathcal{L}_{PC} = \bar{\psi}(i\gamma^\mu\partial_{\mu}-M)\psi 
+ \dfrac{1}{2}{G'_s}^2(\bar{\psi}\psi)^2 - \dfrac{1}{2}{G'_V}^2(\bar{\psi}
\gamma^{\mu}\psi)^2 , 
\label{dlcontato2}
\end{equation}
presenting no mesonic fields.

Let us remark here that the decoupling procedure is just the integration off of 
the mesonic fields leading to contributions in the Lagrangian density, Eq. 
(\ref{dlcontato2}), that are quadratic in the fermion density and fermion 
vector current. Indeed Eq. (\ref{auxiliary}) expresses the auxiliary 
character of the mesonic fields. Their contribution, in the zero order 
approximation, is resumed in the quadratic terms, and no true quanta can be 
assigned to the mesons in this approximation. We have obtained thus a rigorous 
derivation of the equivalence between the linear point-coupling model and the 
zero-order approximation to the Walecka model. This means an equivalence
between the fermion correlation functions, and Wightman functions, of the
models. A reconstruction of the Hilbert space of the zero-order Walecka and
linear point-coupling models will lead to an isomorphism of their Hilbert
spaces. On the other hand, this implies a rigorous derivation of the equivalence
between the mean-field Walecka model and the linear point-coupling one. It means
that both models will lead to the same equations of state as has been pointed
out in \cite{delfino3}.
From the physical point of view it should be said that the meson degrees
of freedom are not excited in the infinite mass limit. This is the physical
mechanism that leads to the isomorphism of the Hilbert spaces of the
models that were pointed here.

\subsection{NLPC model from the MNLW ones}

In the previous section we have shown that the point-coupling model was 
obtained from the Walecka one in the limit of hypermassive mesons. 
To improve the experimental finite nuclei \cite{hussein} and infinite
nuclear matter bulk properties, the well known nonlinear Walecka model
\cite{boguta} adds cubic and quartic scalar self-coupling to the Walecka model, 
\begin{eqnarray}
\mathcal{L}_{NLW} = \mathcal{L}_W - \dfrac{A}{3}\phi^3 - \dfrac{B}{4}\phi^4.
\label{dlwnl}
\end{eqnarray}
Indeed, there is a family of acceptable NLW models which differ in respect to 
how the $A$ and $B$ free parameters are chosen to fit different experimental 
nuclear data \cite{ring}. As we have pointed out before, higher order
point-coupling models involving $(\bar{\psi}\psi)^3$ and $(\bar{\psi}\psi)^4$
(NLPC) have been also successfully applied to finite nuclei \cite{nikolaus}.

Still at the finite range level, i.e., finite meson masses, different
kinds of Walecka-type models such as variants of NLW ones
\cite{variants1,variants2}, with density dependent coupling constants
\cite{ddm}, and the linear chiral model \cite{nav}, were also used in the
description of nuclear frameworks. The NLW models derived from a quark model
perspective can be found in Ref. \cite{tobias}. Particularly, the authors show
that the Walecka model is the limit of infinite quark mass, where the
quarkdynamics freezes.

The question we pose in this section is whether a NLW model, Eq.
(\ref{dlwnl}), leads to a nonlinear point-coupling model in the limit of
infinite meson masses, which includes cubic and quartic self-fermionic terms, 
\begin{eqnarray}
\mathcal{L}_{NLPC} = \bar{\psi}(i\gamma^\mu\partial_{\mu}-M)\psi + 
\dfrac{1}{2}{G'_s}^2(\bar{\psi}\psi)^2 
- \dfrac{1}{2}{G'_V}^2(\bar{\psi}\gamma^{\mu}\psi)^2 
+ \dfrac{A'}{3}(\bar{\psi}\psi)^3 + \dfrac{B'}{4}(\bar{\psi}\psi)^4.
\label{lagrangian34}
\end{eqnarray}
The answer is not. Indeed reproducing the procedure of the last section
instead of equation (\ref{fpwaleckam2}) with (\ref{sfprime}) we obtain,
integrating away the vectorial field
\begin{equation}
W_{MF-NL}[\eta,\bar{\eta}]=N \int[D\psi][D\bar{\psi}][D\phi']\ 
\mbox{e}^{iS'_{S-NL}}
\label{fpwaleckam2nonlinear}
\end{equation}
with
\begin{eqnarray}\label{sfprimenonlinear}
S'_{S-NL} &=& \int d^4x\ \left[\bar{\psi}(i\gamma^\mu\partial_{\mu}-M)\psi 
-\dfrac{1}{2}(\phi'+G'_s \bar{\psi}\psi)^2 + \dfrac{1}{2}{G'_s}^2(\bar{\psi}
\psi)^2 \right.
\nonumber \\ 
&+& \left.  - \dfrac{A}{3}{\phi^\prime}^3 - \dfrac{B}{4}{\phi^\prime}^4- 
\dfrac{1}{2}{G'_V}^2(\bar{\psi}\gamma^{\mu}\psi)^2 + \bar{\eta}(x)\psi(x) 
+ \eta(x)\bar{\psi}(x) \right] \mbox{.}
\end{eqnarray}

Now the functional integral for the field $\phi^\prime$ can not be explicitly 
performed and all we can say is that it gives rises to an unknown functional of 
$\bar\psi\psi$. With this the identification between NLW and NLPC fails. That 
is we can not assert the formal equivalence between NLW and NLPC even at the 
restricted sense of zero-order expansion in the kinetic terms. An approach, 
that connects the NLW and NLPC models, has been nicely performed in
Ref.~\cite{greiner} where the authors use an expansion in the meson propagators 
treating the nonlinearity in the $\phi$ field by an iterative process.

In order to gain a deeper understanding on how to obtain the NLPC model,
Eq. (\ref{lagrangian34}), in the meson hypermassive limit, we consider here a
modification of $\mathcal{L}_{NLW}$ that includes second and third powers of
the scalar meson field coupled to the appropriate powers of the Fermion scalar
density, which allows to decouple the scalar meson field, when the scalar mass
goes to infinity.

We consider then the modified nonlinear Lagrangian, 
\begin{eqnarray}
\mathcal{L}_{MNLW} &=& \mathcal{L}_W  + \mathcal{L}_3 + \mathcal{L}_4
\label{dlmmisto}
\end{eqnarray}
where
\begin{eqnarray}
\mathcal{L}_3 &=& -\dfrac{A'}{3}\left[\left(\dfrac{m_s^2}{g_s}\right)^3\phi^3
+3\left(\dfrac{m_s^2}{g_s}\right)^2\phi^2\bar{\psi}\psi+3\dfrac{m_s^2}{g_s}
\phi(\bar{\psi}\psi)^2\right] 
\label{l3} \quad \mbox{and} \\ 
\mathcal{L}_4 &=& -\dfrac{B'}{4}\left[\left(\dfrac{m_s^2}{g_s}\right)^4\phi^4
+4\left(\dfrac{m_s^2}{g_s}\right)^3\phi^3\bar{\psi}\psi+6\left(\dfrac{m_s^2}
{g_s}\right)^2\phi^2(\bar{\psi}\psi)^2+4\dfrac{m_s^2}{g_s}
\phi(\bar{\psi}\psi)^3\right] \ \mbox{.}
\label{l4}
\end{eqnarray}	

The generating functional for this Lagrangian is the same given by 
Eq. (\ref{fpwalecka}) with $\mathcal{L}_W $ substituted by $\mathcal{L}_{MNLW}$.

By using again the definitions for $V'^{\mu}$, $\phi '$, $G'_s$, $G'_V$, 
$A'_{\mu}$ and $J'$, we see that Eq. (\ref{dlmmisto}) becomes 
\begin{eqnarray}
\mathcal{L}_{MNLW} &=& \bar{\psi}(i\gamma^\mu\partial_{\mu}-M)\psi + \nonumber 
\dfrac{1}{2m_s^2}\partial^{\mu}\phi'\partial_{\mu}\phi' - \dfrac{1}{2}\phi'^2 
-  \dfrac{1}{4m_V^2}F'^{\mu\nu}F'_{\mu\nu} + \dfrac{1}{2}V'_{\mu} V'^{\mu} 
- G'_s \bar{\psi}\phi'\psi  \nonumber \\
&-& G'_V \bar{\psi}\gamma^{\mu}V'_{\mu}\psi + \mathcal{L'}_3 + \mathcal{L'}_4 
\label{dlmmistom} \\
&\equiv& \mathcal{L}'_{MNLW} + \mathcal{L'}_3 + \mathcal{L'}_4 + U(\phi',V'^\mu)
\end{eqnarray}
where
\begin{eqnarray}
\mathcal{L}'_{MNLW} &=& \bar{\psi}(i\gamma^\mu\partial_{\mu}-M)\psi 
- \dfrac{1}{2}\phi'^2  + \dfrac{1}{2}V'_{\mu} V'^{\mu}- G'_s \bar{\psi}\phi'\psi 
- G'_V \bar{\psi}\gamma^{\mu}V'_{\mu}\psi\,\mbox{,} \\
\mathcal{L'}_3 &=& -\dfrac{A'}{3}\left[\dfrac{1}{{G'_s}^3}\phi'^3+\dfrac{3}
{{G'_s}^2}\phi'^2\bar{\psi}\psi+\dfrac{3}{G'_s}\phi'(\bar{\psi}\psi)^2\right]\,
\mbox{,} \\ 
\mathcal{L'}_4 &=& -\dfrac{B'}{4}\left[\dfrac{1}{{G'_s}^4}\phi'^4 
+ \dfrac{4}{{G'_s}^3}\phi'^3\bar{\psi}\psi +\dfrac{6}{{G'_s}^2}\phi'^2
(\bar{\psi}\psi)^2 +\dfrac{4}{G'_s}\phi'(\bar{\psi}\psi)^3\right]
\end{eqnarray}
with  $U(\phi',V'^\mu)$ given by Eq. (\ref{cinetico}). Therefore, the 
generating functional can be rewritten, 
\begin{equation}
W[J',A'_{\mu},\eta,\bar{\eta}]=N \int[D\psi][D\bar{\psi}][DV'^{\mu}][D\phi']
\displaystyle\ \mbox{e}^{\displaystyle i\left[\int d^4x\ U(\phi',V'^\mu) + S'_S\right]}
\label{fpmmistom}
\end{equation}
where 
\begin{equation}
S'_S = \int d^4x\ \left[\mathcal{L}'_{MNLW} + \mathcal{L}'_3 + \mathcal{L}'_4 
+ A'_{\mu}(x)V'^{\mu}(x) + J'(x)\phi'(x) + \bar{\eta}(x)\psi(x) + 
\eta(x)\bar{\psi}(x)\right] \ \mbox{.}
\label{acaommistofontem}
\end{equation}
Again, $U(\phi',V'^\mu)$  will be treated 
perturbatively in the generating functional. The zero order approximation, 
Eq. (\ref{aproximacao}), leads to
\begin{eqnarray}
W[J',A'_{\mu},\eta,\bar{\eta}]=N \int[D\psi][D\bar{\psi}][DV'^{\mu}][D\phi']\ 
\mbox{e}^{iS'_S} \ \mbox{.}
\label{fpmmistom2}
\end{eqnarray}

As we have done previously, we will discard the control on the mesonic 
correlation functions by taking $J'(x)=A'_\mu(x)=0$ in Eq. 
(\ref{acaommistofontem}). Now, together with  Eqs. (\ref{ident1})-(\ref{ident2})
we will use the following set of identities,
\begin{equation}
-\dfrac{A'}{3}\left[\dfrac{1}{{G'_s}^3}\phi'^3+\dfrac{3}{{G'_s}^2}\phi'^2
\bar{\psi}\psi+\dfrac{3}{G'_s}\phi'(\bar{\psi}\psi)^2\right] = -\dfrac{A'}
{3{G'_s}^3}(\phi'+G'_s \bar{\psi}\psi)^3 + \dfrac{A'}{3}(\bar{\psi}\psi)^3 \, 
\mbox{,}
\end{equation}
\vspace{-0.5cm}
\begin{eqnarray}
-\dfrac{B'}{4}\left[\dfrac{1}{{G'_s}^4}\phi'^4 + \dfrac{4}{{G'_s}^3}\phi'^3
\bar{\psi}\psi +\dfrac{6}{{G'_s}^2}\phi'^2(\bar{\psi}\psi)^2 +\dfrac{4}{G'_s}
\phi'(\bar{\psi}\psi)^3\right] &=& -\dfrac{B'}{4{G'_s}^4}(\phi'+G'_s \bar{\psi}
\psi)^4  \nonumber \\
&+& \dfrac{B'}{4}(\bar{\psi}\psi)^4, 
\end{eqnarray}
which allows us rewrite Eq. (\ref{acaommistofontem}) as 
\begin{eqnarray}
S'_S &=& \int d^4x\ \left[\bar{\psi}(i\gamma^\mu\partial_{\mu}-M)\psi 
-\dfrac{1}{2}(\phi'+G'_s \bar{\psi}\psi)^2 + \dfrac{1}{2}{G'_s}^2(\bar{\psi}
\psi)^2 \right.
\nonumber \\ 
&+& \dfrac{1}{2}(V'^\mu - G'_V \bar{\psi}\gamma^{\mu}\psi)^2 
- \dfrac{1}{2}{G'_V}^2(\bar{\psi}\gamma^{\mu}\psi)^2 -\dfrac{A'}{3{G'_s}^3}
(\phi'+G'_s \bar{\psi}\psi)^3 + \dfrac{A'}{3}(\bar{\psi}\psi)^3 
\nonumber \\
&-& \left. \dfrac{B'}{4{G'_s}^4}(\phi'+G'_s \bar{\psi}\psi)^4 + \dfrac{B'}{4}
(\bar{\psi}\psi)^4 + \bar{\eta}(x)\psi(x) + \eta(x)\bar{\psi}(x) \right] 
\mbox{.}
\end{eqnarray}

If we define once again the change of fields leading to the auxiliary fields,
\begin{eqnarray}
\lambda &=& \phi' + G'_s \bar{\psi}\psi \\
R^\mu &=&  V'^\mu - G'_V \bar{\psi}\gamma^{\mu}\psi, 
\end{eqnarray}
we will have for the mesonic field integrals in Eq. (\ref{fpmmistom}) 
the following forms, 
\begin{equation} \nonumber
\int[D\phi']\ \mbox{e}^{\displaystyle -i \int d^4x \ 
\left[\dfrac{1}{2}(\phi'+G'_s\bar{\psi}\psi)^2 + \dfrac{A'}{3{G'_s}^3}
(\phi'+G'_s \bar{\psi}\psi)^3 + \dfrac{B'}{4{G'_s}^4}
(\phi'+G'_s \bar{\psi}\psi)^4 \right]} =
\end{equation}
\vspace{-0.8cm}
\begin{equation}
= \int[D\lambda]\ \mbox{e}^{\displaystyle -i\int d^4x \ 
\left[ \dfrac{1}{2}\lambda^2 + \dfrac{A'}{3{G'_s}^3}\lambda^3 
+ \dfrac{B'}{4{G'_s}^4}\lambda^4 \right]}
\end{equation}
and 
\vspace{0.2cm}
\begin{equation}
\int[DV'^\mu]\ \mbox{e}^{\displaystyle i \int d^4x\ \dfrac{1}{2}
(V'^\mu-G'_V\bar{\psi}\gamma^{\mu}\psi)^2} = \int[DR^\mu]\ 
\mbox{e}^{\displaystyle i \int d^4x \ \dfrac{1}{2}R^\mu R_\mu} \mbox{.}
\end{equation}
The identities and translations above allows to rewrite the Eq. 
(\ref{fpmmistom2}) as 
\begin{equation}
W[\eta,\bar{\eta}]=\mathcal{N} \int[D\psi][D\bar{\psi}]\ \mbox{e}^{i S''_S}
\end{equation}
where 
\begin{eqnarray}
{\mathcal{N}}^{-1} &=& {\displaystyle \int[D\psi][D\bar{\psi}]\ 
\mbox{e}^{i S''}}\,\mbox{,} \\
S'' &=& \int d^4x\ \left[\bar{\psi}(i\gamma^\mu\partial_{\mu}-M)\psi 
+ \dfrac{1}{2}{G'_s}^2(\bar{\psi}\psi)^2 - \dfrac{1}{2}{G'_V}^2
(\bar{\psi}\gamma^{\mu}\psi)^2 + \dfrac{A'}{3}(\bar{\psi}\psi)^3 
+ \dfrac{B'}{4}(\bar{\psi}\psi)^4 \right]
\label{acaocontato34} \quad \,\, \\
S''_S &=& \int d^4x \ \left[\bar{\psi}(i\gamma^\mu\partial_{\mu}-M)\psi 
+ \dfrac{1}{2}{G'_s}^2(\bar{\psi}\psi)^2 
- \dfrac{1}{2}{G'_V}^2(\bar{\psi}\gamma^{\mu}\psi)^2 
+ \dfrac{A'}{3}(\bar{\psi}\psi)^3 \right. \nonumber \\
&+& \left. \dfrac{B'}{4}(\bar{\psi}\psi)^4 + \bar{\eta}(x)\psi(x) 
+ \eta(x)\bar{\psi}(x) \right] \ \mbox{.}
\label{acaofontecontato34}
\end{eqnarray}

The Lagrangian density contained in Eqs.
(\ref{acaocontato34})-(\ref{acaofontecontato34}) describes the fermionic
nonlinear point-coupling 
model we have aimed. We have seen  that the generating  Lagrangian to obtain 
the NLPC model through the mesonic hypermassive limit is  $\mathcal{L}_{MNLW}$, 
given by Eq. (\ref{dlmmisto}) and not  $\mathcal{L}_{NLW}$ as could be naively 
expected. 

Let us perform here an analysis based upon the structural properties of the
model. We are not asserting here an equivalence of NLPC and MNLW models. The
Hilbert spaces of both models are not isomorphic. But the zero order 
expansion of the MNLW model has been exactly mapped onto the NLPC models.
Once again from the view point of an structural analysis the irreducible
algebra of fields of MNLW composed of the polynomial algebra of
 the fermion and meson fields becomes reducible
in the zero order approximation. The mesonic fields turn out to be functions
of the fermion fields and the irreducible algebra is composed solely of the
fermion field algebra.
In the language of functional integrals this is implemented
by the decoupling of the auxiliary fields $\lambda$ and $R^\mu$. The equations
of motion of the auxiliary fields bring about the functional relation between 
the original mesonic fields and the fermion bilinears. Contrary to the linear 
case treated in the preceding section now the equations of motion do not demand 
$\lambda=0$ and $R^\mu =0$. The equations for the auxiliary fields includes, in 
principle, other roots besides the trivial ones. Actually, as we will
discuss in the next section, the equations of state of the MNLW model in the
mean-field approximation, depend on the mean value of the auxiliary field
$\lambda$ and differ from those of the NLPC model only by the terms containing
this field. However, the physical requirement of vanishing pressure at zero
Fermi momentum is satisfied only by the trivial solution for the auxiliary
field $\lambda$.

Another aspect that should be stressed is concerned to   the renormalization
properties of the models. The infinite mass expansion that is done here
effectively changes the power counting dimensions of the mesonic fields. Since
their kinetic terms are discarded, there appear no inverse powers of the momenta
in their propagators in the ultraviolet region. The result is that the ${\cal
L}_{NLPC}$  models are non-renormalizable while the ${\cal L}_{NLW}$ are
(power-counting) renormalizable. The physical reason for this change shall be
emphasized: our approximation freezes the meson degrees of freedom that are
necessary to render the Walecka model (power-counting) renormalizable.  

\section{MEAN-FIELD APPROXIMATION} 

Now we perform an alternative procedure to derive the NLPC model from the 
meson-exchange MNLW one. Here we perform the largely used mean-field
approach (MFA), instead of the infinite meson mass limit taken in the previous 
section. We also use the no-sea approximation, i.e., we consider only
the valence Fermi states. We will show that the EOS of the MNLW model are
exactly the same of the NLPC one. 

For infinite nuclear matter the energy density, and pressure of the NLPC models
are given respectively by
\begin{eqnarray}
\mathcal{E} = \dfrac{1}{2}{G'_V}^2\rho^2 + \dfrac{1}{2}{G'_s}^2\rho_s^2 
+ \dfrac{2}{3}A'\rho_s^3 + \dfrac{3}{4}B'\rho_s^4 + \dfrac{\gamma}{2\pi^2} 
\int_0^{k_F}k^2(k^2 + {M^*}^2)^{1/2}dk
\label{energy}
\end{eqnarray}
and
\begin{eqnarray}
P = \dfrac{1}{2}{G'_V}^2\rho^2 - \dfrac{1}{2}{G'_s}^2\rho_s^2 
- \dfrac{2}{3}A'\rho_s^3 - \dfrac{3}{4}B'\rho_s^4 
+ \dfrac{\gamma}{6\pi^2}\int_0^{k_F}\dfrac{k^4}{(k^2 + {M^*}^2)^{1/2}},
\label{press}
\end{eqnarray}
with the vector, and the scalar density defined as 
\begin{eqnarray}
\rho &=&\dfrac{\gamma}{2\pi^2}\int_0^{k_F}k^2dk \quad\mbox{and} \\
\rho_s&=&\dfrac{\gamma}{2\pi^2}\int_0^{k_F}\frac{M^*}{(k^2+{M^*}^2)^{1/2}}k^2dk,
\label{rhoconttemp}
\end{eqnarray}
with $k_F$ being the Fermi momentum, $\gamma=4$ for symmetric nuclear matter and
$\gamma=2$ for neutron matter. The nucleon effective mass reads
\begin{equation}
M^* \equiv M - {G'_s}^2\rho_s  - A'\rho_s^2 - B'\rho_s^3.
\label{mefcontato}
\end{equation}

Let us now start with the derivation of the MNLW equations of state first 
rewriting its Lagrangian, Eq. (\ref{dlmmisto}), as follows
\begin{eqnarray}
\mathcal{L}_{MNLW} &=& \bar{\psi}\left( i\gamma^\mu\partial_\mu-M\right)\psi 
+ \frac{1}{2}\partial^\mu \phi \partial_\mu \phi 
- \frac{1}{4}F^{\mu\nu}F_{\mu\nu} + \frac{1}{2}m^2_V V_\mu V^\mu 
- g_V\bar{\psi}\gamma^\mu V_ \mu\psi\nonumber \\
&-& \frac{1}{2}\left( m_s\phi + \frac{g_s}{m_s}\bar{\psi}\psi \right)^2 
+ \frac{g_s^2}{2m^2_s}\left(\bar{\psi}\psi\right)^2 
- \frac{A'}{3}\left(\frac{m_s}{g_s}\phi
+ \bar{\psi}\psi\right)^3 
+ \frac{A'}{3}\left(\bar{\psi}\psi\right)^3\nonumber \\
&-& \frac{B'}{4}\left(\frac{m^2_s}{g_s}\phi + \bar{\psi}\psi\right)^4 
+ \frac{B'}{4}\left(\bar{\psi}\psi\right)^4 \, \mbox{.}
\label{lm}
\end{eqnarray}
By knowing that a field translation does not alter the physical content of the 
model, we define 
\begin{eqnarray}
\lambda \equiv \frac{m^2_s}{g_s}\phi + \bar{\psi}\psi \, \mbox{.}
\end{eqnarray}
With this definition, the MNLW Lagrangian acquires the following form, 
\begin{eqnarray}
\mathcal{L}_{MNLW} &=& \bar{\psi}\left( i\gamma^\mu\partial_\mu-M\right)\psi  
- \frac{1}{4}F^{\mu\nu}F_{\mu\nu} + \frac{1}{2}m^2_V V_\mu V^\mu 
- g_V\bar{\psi}\gamma^\mu V_ \mu\psi - \frac{1}{2}{G'_s}^2\lambda^2 \nonumber \\
&+& \dfrac{1}{2}{G'_s}^2\left(\bar{\psi}\psi\right)^2 - \frac{A'}{3}\lambda^3 
+ \frac{A'}{3}\left(\bar{\psi}\psi\right)^3 - \frac{B'}{4}\lambda^4 
+ \frac{B'}{4}\left(\bar{\psi}\psi\right)^4 \nonumber \\
&+& \dfrac{G'_s}{2m_s^2}\left(\partial^\mu\lambda 
- \partial^\mu\bar{\psi}\psi \right)\left(\partial_\mu\lambda 
- \partial_\mu\bar{\psi}\psi \right), 
\label{dlmm}
\end{eqnarray}
where $G'_s = g_s/m_s$.

In the MFA, the scalar and vector mesonic fields are replaced by their average 
values, 
\begin{eqnarray}
\lambda \rightarrow \left\langle \lambda \right\rangle &\equiv& \lambda \\
V^\mu \rightarrow \left\langle V^\mu \right\rangle &\equiv& \delta^{\mu 0}V^0 \, 
\mbox{.}
\end{eqnarray}
Still in this approximation, we use the $\bar{\psi}\psi$ ground-state
expectation value. It is also assumed that the system is spatially uniform, so
that the derivative terms of  $\lambda$ and $\bar{\psi}\psi$ disappear. 
Therefore, Eq. (\ref{dlmm}) becomes,
\begin{eqnarray}
\mathcal{L}_{MNLW}^{(MFA)}&=&\bar{\psi}\left( i\gamma^\mu\partial_\mu-M\right)
\psi + \frac{1}{2}m^2_V V^2_0 - g_V\bar{\psi}\gamma^0 V_0\psi 
- \frac{1}{2}{G'_s}^2\lambda^2 + \frac{1}{2}{G'_s}^2
\left(\bar{\psi}\psi\right)^2 \nonumber \\
&-& \frac{A'}{3}\lambda^3 + \frac{A'}{3}\left(\bar{\psi}\psi\right)^3 
- \frac{B'}{4}\lambda^4 + \frac{B'}{4}\left(\bar{\psi}\psi\right)^4 \, \mbox{.}
\end{eqnarray}

The independent fields of this theory may be taken as $\lambda$, $\bar{\psi}$,
$\psi$ and $V^0$. From the Euler-Lagrange equations one obtains the equations of
motion for the fields,
\begin{equation}
\lambda({G'_s}^2 + A'\lambda + B'\lambda^2) = 0 \, \mbox{,}
\label{emlambda}
\end{equation}
\begin{equation}
V_0 = \frac{g_V}{m^2_V}\bar{\psi}\gamma_0\psi
\label{emv0}
\end{equation}
and 
\begin{equation}
[ i\gamma^\mu\partial_\mu - g_V\gamma^0V_ 0 - (M - {G'_s}^2(\bar{\psi}\psi) 
- A'(\bar{\psi}\psi)^2 - B'(\bar{\psi}\psi)^3)]\psi =0 \, \mbox{.}
\label{empsi} 
\end{equation}
Now, substituting  $\bar{\psi}\psi$ and $\bar{\psi}\gamma_{\mu}\psi$ 
by their respective mean values, we have 
\begin{eqnarray}
V_0 = \frac{g_V}{m^2_V} \left< \bar{\psi}\gamma_0\psi \right> 
= \frac{g_V}{m^2_V}\rho
\label{v0}
\end{eqnarray}
and consequently, Eq. (\ref{empsi}) may be rewritten as 
\begin{eqnarray}
[ i\gamma^{\mu}\partial_{\mu} - \gamma^0{G'_V}^2\rho - (M - {G'_s}^2\rho_s  
- A'\rho_s^2 - B'\rho_s^3)]\psi=0 , 
\end{eqnarray}
where ${G'}_V^2 = g_V^2/m_V^2$, ${G'}_s^2 = g_s^2/m_s^2$, and $\rho_s 
= \left<\bar{\psi}\psi\right>$.
The above Dirac equation suggests the effective nucleon mass definition, 
\begin{eqnarray}
M^* = M - {G'_s}^2\rho_s - A'\rho^2_s - B'\rho^3_s,
\end{eqnarray}
that is exactly the same expression given in Eq. (\ref{mefcontato}).

The mean-field equations of state will come from the energy-momentum 
tensor, 
\begin{eqnarray}
T^{(MFA)}_{\mu\nu} &=& -g_{\mu\nu} \left[ \bar{\psi}
(i\gamma^\alpha\partial_\alpha - g_V\gamma^0 V_0 - M)\psi 
+ \dfrac{1}{2}m^2_V V^2_0 - \dfrac{1}{2}{G'_s}^2\lambda^2 
+ \dfrac{1}{2}{G'_s}^2(\bar{\psi}\psi)^2\right. \nonumber \\ 
&-& \left. \dfrac{A'}{3}\lambda^3 + \dfrac{A'}{3}(\bar{\psi}\psi)^3 
- \dfrac{B'}{4}\lambda^4 + \frac{B'}{4}\left(\bar{\psi}\psi\right)^4 \right] 
+ i\bar{\psi}\gamma_\mu\partial_\nu\psi \nonumber \\
&=& -g_{\mu\nu} \left[ - \dfrac{1}{2}{G'_s}^2(\bar{\psi}\psi)^2 
- \dfrac{2A'}{3}(\bar{\psi}\psi)^3 - \dfrac{3B'}{4}(\bar{\psi}\psi)^4 
+ \dfrac{1}{2}m^2_V V^2_0 - \dfrac{1}{2}{G'_s}^2\lambda^2 \right. \nonumber \\
&-&\left. \dfrac{A'}{3}\lambda^3 - \dfrac{B'}{4}\lambda^4 \right] 
+ i\bar{\psi}\gamma_\mu\partial_\nu\psi \, \mbox{.}
\end{eqnarray}

The density energy is obtained from 
\begin{eqnarray}
\mathcal{E} &=& \left< T^{(MFA)}_{00} \right> \nonumber \\
&=& \dfrac{1}{2}{G'_s}^2\rho^2_s + \dfrac{2A'}{3}\rho^3_s 
+ \dfrac{3B'}{4}\rho^4_s - \dfrac{1}{2}{G'_V}^2\rho^2 
+ \dfrac{1}{2}{G'_s}^2\lambda^2 + \dfrac{A'}{3}\lambda^3 
+ \dfrac{B'}{4}\lambda^4 + i\left< \bar{\psi}\gamma_0\partial_0\psi \right> , 
\end{eqnarray}
where we have used Eq. (\ref{v0}). 

The quantity $i\left< \bar{\psi}\gamma_0\partial_0\psi \right>$ is found by 
using the dispersion relation $k_0 = g_V V_0 + (k^2 + {M^*}^2)^{1/2} 
= {G'_V}^2 \rho + (k^2 + {M^*}^2)^{1/2}$, where $k_0$ is the fourth 
energy-momentum component. This leads to 
\begin{eqnarray}
\mathcal{E} &=& \dfrac{1}{2}{G'_V}^2\rho^2 + \dfrac{1}{2}{G'_s}^2\rho^2_s 
+ \dfrac{2A'}{3}\rho^3_s + \dfrac{3B'}{4}\rho^4_s 
+ \dfrac{1}{2}{G'_s}^2\lambda^2 + \dfrac{A'}{3}\lambda^3 
+ \dfrac{B'}{4}\lambda^4 \nonumber \\ 
&+& \dfrac{\gamma}{2\pi^2} \int_0^{k_F} (k^2 + {M^*}^2)^{1/2} k^2 dk \ \mbox{.}
\label{elambda}
\end{eqnarray}

The pressure is obtained by 
\begin{eqnarray}
P &=& \dfrac{1}{3}\left< T_{ii}\right> \nonumber \\	
&=& \dfrac{1}{2}{G'_V}^2\rho^2 - \dfrac{1}{2}{G'_s}^2\rho^2_s 
- \dfrac{2A'}{3}\rho^3_s - \dfrac{3B'}{4}\rho^4_s 
- \dfrac{1}{2}{G'_s}^2\lambda^2 - \dfrac{A'}{3}\lambda^3 
- \dfrac{B'}{4}\lambda^4
+ \dfrac{1}{3}i\left\langle \bar{\psi}\gamma_i \partial_i \psi \right\rangle .
\end{eqnarray}

By extracting 
$i\left\langle \bar{\psi}\gamma_i \partial_i \psi \right\rangle$ 
from the the Dirac equation, the pressure can be written as follows
\begin{eqnarray}
P &=& \dfrac{1}{2}{G'_V}^2\rho^2 - \dfrac{1}{2}{G'_s}^2\rho^2_s 
- \dfrac{2A'}{3}\rho^3_s - \dfrac{3B'}{4}\rho^4_s 
- \dfrac{1}{2}{G'_s}^2\lambda^2 - \dfrac{A'}{3}\lambda^3 
- \dfrac{B'}{4}\lambda^4 \nonumber \\ 
  &+& \dfrac{\gamma}{6\pi^2}\int_0^{k_F}\dfrac{k^4}{(k^2 + {M^*}^2)^{1/2}}dk 
\ \mbox{.}
\label{plambda}
\end{eqnarray}
The auxiliary $\lambda$ field are decoupled from the fermionic sector. Its
contribution either to the pressure or to the energy shall indeed be dropped by
the physical requirement that the pressure goes to zero when $k_F$ vanishes,
implying that only the trivial solution, $\lambda=0$, of Eq.
(\ref{emlambda}) should be kept. Therefore the energy density and the pressure
become, respectively
\begin{equation}
\mathcal{E} = \dfrac{1}{2}{G'_V}^2\rho^2 + \dfrac{1}{2}{G'_s}^2\rho^2_s 
+ \dfrac{2}{3}A'\rho^3_s + \dfrac{3}{4}B'\rho^4_s + \dfrac{\gamma}{2\pi^2} 
\int_0^{k_F} (k^2 + {M^*}^2)^{1/2} k^2 dk
\end{equation}
and
\begin{equation}
P = \dfrac{1}{2}{G'_V}^2\rho^2 - \dfrac{1}{2}{G'_s}^2\rho^2_s 
- \dfrac{2}{3}A'\rho^3_s - \dfrac{3}{4}B'\rho^4_s 
+ \dfrac{\gamma}{6\pi^2}\int_0^{k_F}\dfrac{k^4}{(k^2 + {M^*}^2)^{1/2}}dk. 
\end{equation}

Notice that the equations above of the MNLW model are identical to Eqs.
(\ref{energy}), and (\ref{press}) of the NLPC ones, what shows that the former
can be also obtained from the mean-field approximation at the level of the EOS
instead of the Lagrangian density framework shown by the hypermassive limit
taken in the Section II.

\section{CONCLUSION}  

If one performs a hypermassive meson limit to usual NLW models, baryonic NLPC
models are not obtained. We have shown that in order to obtain NLPC models, a
modification of NLW models are needed, already at the level of the Lagrangian
density. In this work we derived the point-coupling models from a modified NLW
by using the hypermassive meson limit in the functional integral method. From
this approach we shown how the linear PC models can be obtained from the Walecka
ones, and in the same way, how the MNLW model generates the NLPC one. This
relation among MNLW and NLPC is described as equivalences of the physical
content of these models encoded in their irreducible algebra of fields in the
infinite meson masses limit. In addition, we also use the no-sea approximation
to construct the equations of state of the MNLW model. We remark that even in
this alternative way (without taking the hypermassive meson limit) the EOS are
exactly the same of the NLPC ones. Therefore, to treat MNLW model either from
the meson hypermassive limit or alternatively to treat directly from the mean
field approach leads to the same NLPC models.

\section*{ACKNOWLEDGEMENTS}

This work was supported by the Brazilian agencies FAPERJ, CAPES (Proc. BEX 0885/11-8), CNPq, and
FAPESP.

\end{document}